# Emulated ASIC Power and Temperature Monitor System for FPGA Prototyping of an Invasive MPSoC Computing Architecture

Elisabeth Glocker\*, Qingqing Chen†, Asheque M. Zaidi\*, Ulf Schlichtmann† and Doris Schmitt-Landsiedel\*
\*Institute for Technical Electronics †Institute for Electronic Design Automation
Technische Universität München, Munich, Germany
Email: elisabeth.glocker@tum.de

*Abstract*—In this contribution the emulation of an ASIC temperature and power monitoring system (TPMon) for FPGA prototyping is presented and tested to control processor temperatures under different control targets and operating strategies. The approach for emulating the power monitor is based on an instruction-level energy model. For emulating the temperature monitor, a thermal RC model is used. The monitoring system supplies an invasive MPSoC computing architecture with hardware status information (power and temperature data of the processors within the system). These data are required for resource-aware load distribution. As a proof of concept different operating strategies and control targets were evaluated for a 2-tile invasive MPSoC computing system.

## I. Introduction

The demand for increasing performance and at the same time decreasing feature sizes in modern CMOS technologies lead to higher power densities that increase processor temperature. High temperature reduces circuit speed and causes transistor degradation. If the workload is distributed in a non-uniform way in different circuit blocks, which is typically the case for processors, local temperature hot spots arise. This can greatly affect the lifetime and reliability of a chip [3], [8].

For multi-processor systems on chip (MPSoCs) this becomes even more critical, since a core's temperature depends not only on its own power density, but also on the activity and power density of neighbour cores. Also temperatures are distributed non-uniformly not only within one core, but also from one core to another. Since the processing load for processors may vary extremely, this results in significant differences in processing capability and vulnerability to degradation. Even processors of the same type may behave very differently because of those differences and hardware imperfections. This might also change during the lifetime of the system.

By monitoring the operating temperature and power during run-time, it is possible to control and limit them by load balancing to increase lifetime and reliability and also find best suitable application-processor pairs during resource allocation which becomes especially important for processors with varying processing capabilities. Monitoring of both power and temperature is needed to consider current core-local hardware status provided by the power monitor and also inter-core hardware status provided by the temperature monitor, since temperature is influenced by neighbour core activities.

Invasive computing [10] is a resource-aware programming paradigm where resources (e.g. processors) are temporarily allocated to an application according to its temporal needs dynamically during run-time in a self-organising manner. Current physical hardware properties like power, temperature and aging should be included during resource allocation to select best suitable resource-application pairs and reach system targets (e.g. limit temperature to increase reliability).

FPGA prototyping is often required for design phase verification and design parameter tuning before the envisioned ASIC is produced, where real temperature sensors are used. Monitor modules with analog-circuit implementation, like power and temperature monitors, cannot directly be implemented on the prefabricated FPGA structure. Also an FPGA will show very different behaviour compared to the envisioned ASIC implementation. To nevertheless evaluate the performance of resource allocation using monitoring data, the behaviour of an ASIC monitoring system was emulated on FPGA. For reliable evaluation, a realtime monitor system useable during run-time is needed that runs at the same speed as the processor prototype and delivers up-to-date data. There are approaches for power monitors (e.g. [4]) that could still be much slower than a processor prototype. Temperature analyses in a running system rely on real temperature sensors or on a host PC running thermal modelling tools as in [1] that could also be much slower than a processor prototype.

## II. TPMon Implementation and Usage Strategies

The implemented TPMon (Fig. 1) emulates the behaviour of an ASIC monitor system for an invasive MPSoC architecture with LEON3 cores in TSMC 90nm technology with a clock frequency of 400 MHz. Each tile contains one TPMon (containing a single temperature monitor and for every core a power monitor). Emulation of the power monitor is based on an energy model on instruction level [2], [7], stored in a Look-Up-Table (Power LUT) that contains the average energy consumption for all possible instructions of the LEON core. Emulation of the temperature monitor is based on a thermal RC model (details in [5], [9]) that considers a core's current temperature (Temp. LUT) and the influence on this temperature based on neighbour core activity (Neighbour effect). The maximal delay from monitor input to output was found to be less than $12\,\text{ns}$. So TPMon can be used with clock frequencies of up to about $80\,\text{MHz}$ and is suitable for the invasive FPGA prototype that has a maximal clock frequency of $50\,\text{MHz}$. TPMon accumulates the power of the individual processors. After a time step of $1\,\mu\text{s}$ ($400$ clock cycles for ASIC-core with clock frequency of $400\,\text{MHz}$) the corresponding temperature values are determined. Accumulated power and temperature values are read out and used for tile-local resource allocation. Further accumulated and abstracted values are given to the run-time support system for inter-tile resource allocation.

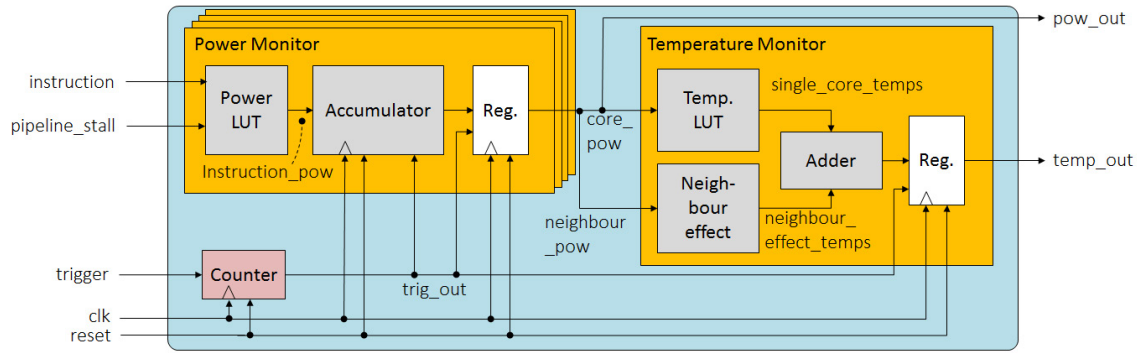

Fig. 1. Implementation of the TPMon for tiles with 2x2 cores. One power monitor is needed for each core, so it is duplicated four times.

Monitor data can be used to control temperature with intelligent load balancing and resource allocation (e.g. outbalance high temperatures in the past with low ones in the present) under different control targets and to deliver monitor data for resource allocation. Different operating strategies and control targets were evaluated for a 2-tile invasive MPSoC computing system (each tile with 2x2 cores) by using TPMon as proof of concept: As example eight tasks (four result in lowest-power consumption and the other four result in highest power consumption) are assumed that should be mapped to the cores. Different resource allocation strategies of neighbour cores can lead to different temperature evolutions, for the same task running on a core. If only one core in a tile with medium-power consumption is active, the resulting temperature is $47\,°\text{C}$. If all neighbour cores of the tile are also active with highest power consumption, this results in a temperature of $53\,°\text{C}$ (+13%). If the control target is to keep the temperatures as low as possible, a strategy to achieve lowest global maximum temperature should be chosen (in the example: a strategy with 2 cores with highest and 2 cores with lowest power consumption per tile. This results in a max. temperature of $51\,°\text{C}$ in both tiles). The lower the core temperature, the better is also the reduction of aging. Thus core's lifetime and reliability increase. If the control target is to keep the temperature difference between cores inside a tile low (to ensure similar core aging inside the tile), a strategy to achieve the same core temperatures within one tile should be chosen (in the example: a strategy with cores with highest power consumption on one tile and the ones with lowest power consumption on the other. This results in one tile with core temperatures of $54\,°\text{C}$ and one with $47\,°\text{C}$).

## III. CONCLUSION AND FUTURE WORK

With the presented FPGA emulation of an ASIC power and temperature run-time monitoring system, an invasive MPSoC computing architecture can be supplied with hardware-status information—in terms of power and temperature data of the processors. This monitor emulation is required to evaluate the entire MPSoC system during the FPGA prototyping stage, before the system is then realized as an ASIC on which the temperature monitors will be analog circuits.

As future work, we will optimize the monitor system within the whole invasive MPSoC computing architecture by a detailed study of the resource allocation process taking monitoring data into account. We will also emulate aging monitors by using aging models [6].

### ACKNOWLEDGMENT

This work was supported by the German Research Foundation (DFG) as part of the Transregional Collaborative Research Centre "Invasive Computing" (SFB/TR 89).